\documentclass{epl}
\title{Comment on ``Equivalence of the variational matrix product method
	and the density matrix renormalization group applied to spin chains''
	by J. Dukelsky, M. A. Mart\'in-Delgado, T. Nishino and G. Sierra}
\shorttitle{Comment on ``Equivalence of the variational matrix product\ldots''}
\author{I. P. McCulloch\inst{1} \and M. Gul\'acsi\inst{2}}
\institute{
  \inst{1} Instituut-Lorentz - Universiteit Leiden, The Netherlands\\
  \inst{2} Department of Theoretical Physics - Australian National University, Australia
}
\begin{document}
\maketitle
\newcommand{\eonesite}{$e^{\mbox{\scriptsize 1 site}}$\ }
\newcommand{\etwosite}{$e^{\mbox{\scriptsize 2 sites}}$\ }
\newcommand{\edmrgjd}{$e^{\mbox{\scriptsize DMRG DMNS}}$\ }
\newcommand{\empjd}{$e^{\mbox{\scriptsize MP DMNS}}$\ }
\newcommand{\vektor}[1]{\mbox{\boldmath $#1$}}
Dukelsky, Mart\'in-Delgado, Nishino and Sierra \cite{Duk} (hereafter referred to as DMNS)
investigated the matrix product method (MPM)\cite{MPM}, comparing
it with the infinite-size density matrix renormalization group (DMRG)\cite{White}.
For equivalent basis size, the MPM produces an improved variational energy over
that produced by DMRG and, unlike DMRG, produces a translationally-invariant wavefunction.
The DMRG results presented were significantly worse than the MPM, caused by
a shallow bound state appearing at the join of the two DMRG blocks. They also suggested
that the DMRG results can be improved by using an alternate superblock construction
$[B] \bullet [B]$ for the last few steps of the calculation.

In this comment, we show that the DMRG results presented by DMNS are in
error and the artificial bound state produced by the standard superblock configuration is very
small even for $m=2$ states kept. In addition, we calculate explicitly the energy and
wavefunction for the $[B] \bullet [B]$ superblock structure and verify that the
energy coincides with that of the MPM, as conjectured in \cite{MPM}.

\begin{figure}[ht]
\center{\includegraphics[width=4.0cm,angle=270]{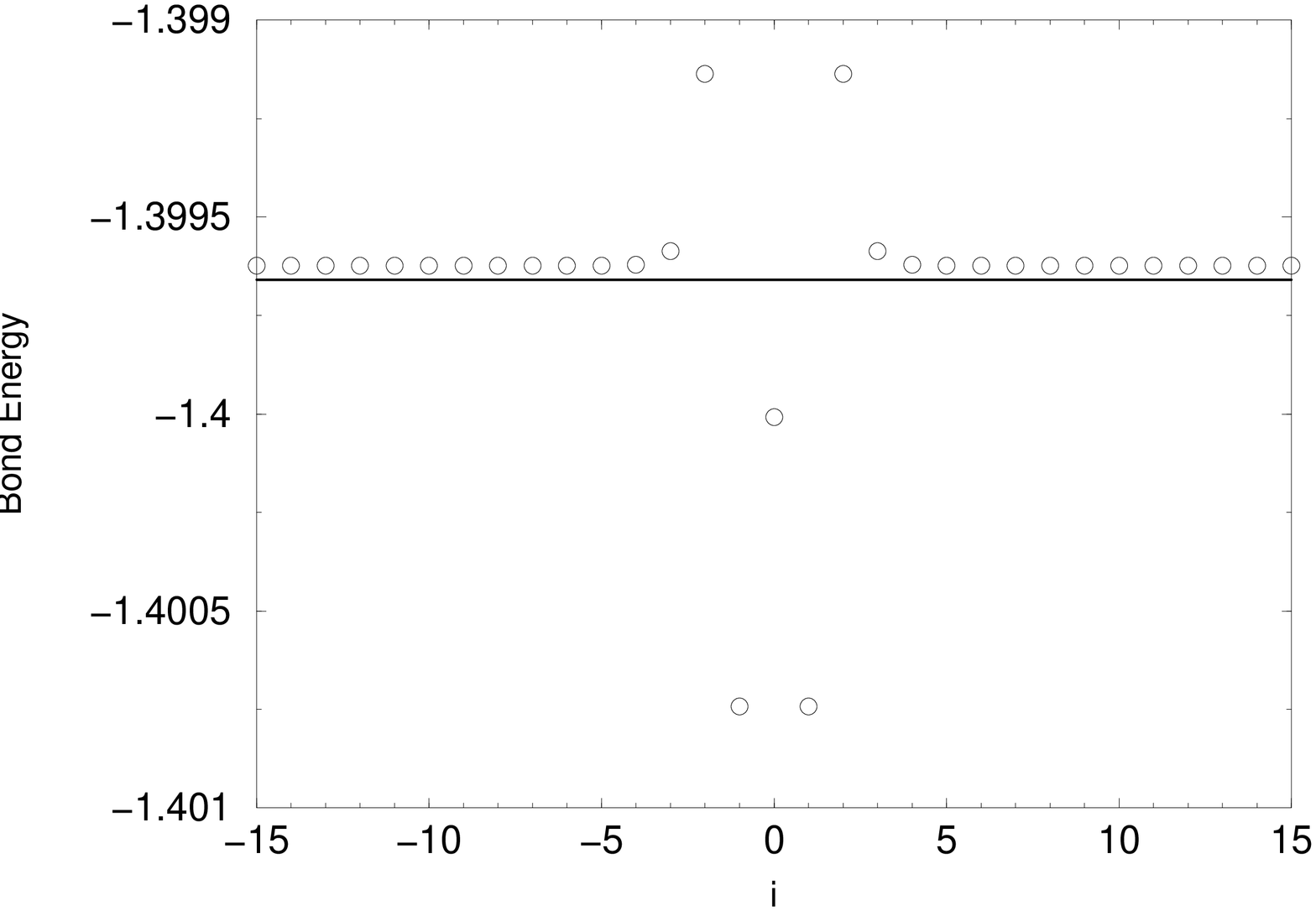}}
\caption{
Bond energy $\left< \vektor{S}_i \cdot \vektor{S}_{i+1} \right>$ as a function of lattice position for the
DMRG wavefunction of the spin 1 Heisenberg chain with $m=2$.
}
\label{fig:BondEnergy}
\end{figure}

The matrix product method allows full $SU(2)$ symmetry to be utilized in a natural
way, and is relatively easy to implement in a practical calculation. On the other hand,
utilizing such non-Abelian symmetries in DMRG appears, at first sight, to be a more difficult
problem and it was not until the invention of the interaction-round-a-face DMRG (IRF-DMRG)
by Sierra and Nishino \cite{IRFDMRG} that the first breakthrough occurred. Recently,
it has been shown that non-Abelian symmetries can be integrated into the DMRG
algorithm directly, without the need for a vertex-IRF transformation\cite{MyThesis,NonAbelian}. This results in
several simplifications over the IRF-DMRG algorithm, while in principle the
numerical results should be identical for the two methods. Table~\ref{tbl:OurDMRGTable}
shows the results of a re-examination of the spin-1 Heisenberg chain using the non-Abelian method,
originally carried out in reference \cite{MyThesis}.
The results of DMNS for the ground state energy determined by
the MPM and the IRF-DMRG are listed as \empjd and \edmrgjd respectively. 
Current results
using the superblock structure $[B] \bullet [B]$ and $[B] \bullet \bullet [B]$ are listed
in columns \eonesite and \etwosite respectively. The DMRG results of 
DMNS are claimed to use the $[B] \bullet \bullet [B]$ structure, hence the
energies in columns \edmrgjd and \etwosite should agree exactly. The discrepancy is,
we believe, due to a problem with the implementation of the IRF-DMRG algorithm used
by DMNS. While there is still a shallow bound state located
in the center of the superblock, the effect on the ground state energy is several
orders of magnitude smaller than reckoned by DMNS. 
Figure \ref{fig:BondEnergy} shows the bond energy $\left< \vektor{S}_i \cdot \vektor{S}_{i+1} \right>$
as a function of lattice position for $m=2$ states kept. For the superblock structure
$[B] \bullet [B]$, the bond energy is exactly translationally invariant as predicted by
DMNS and is shown here with a solid line. In this case,
the bond energy for different values of $m$ (column \eonesite in table~\ref{tbl:OurDMRGTable}) 
agrees with that calculated by
the matrix product method by DMNS (column \empjd in table~\ref{tbl:OurDMRGTable}), 
verifying the conjecture that the wavefunctions
produced by these two algorithms coincide in the thermodynamic limit.
\begin{table}
\caption{Energy density of the spin 1 Heisenberg chain as a function of number of states
	kept. \empjd\ and \edmrgjd\, are from reference \cite{Duk}.
}
\label{tbl:OurDMRGTable}
\[
\begin{array}{rr@{.}lc@{}r@{.}lc@{}r@{.}lc@{}r@{.}lc@{}r@{\times}lr}
m & \multicolumn{2}{c}{\mbox{\empjd}} & & \multicolumn{2}{c}{\mbox{\edmrgjd}} & &
	\multicolumn{2}{c}{\mbox{\eonesite}} & & \multicolumn{2}{c}{\mbox{\etwosite}} & &
	\multicolumn{2}{c}{1 - P_m} & \\
\hline
1 & -1 & 333333 & & -1 & 333333 & & -1 & 3333333 & & -1 & 3333333 & & 1.58 & 10^{-2} & \\
2 & -1 & 399659 & & -1 & 369077 & & -1 & 3996590 & & -1 & 3996237 & & 4.06 & 10^{-4} & \\
3 & -1 & 401093 & & -1 & 392515 & & -1 & 4010933 & & -1 & 4010886 & & 5.39 & 10^{-5} & \\
4 & -1 & 401380 & & -1 & 401380 & & -1 & 4013806 & & -1 & 4013798 & & 1.63 & 10^{-5} & \\
5 & -1 & 401443 & & -1 & 401436 & & -1 & 4014447 & & -1 & 4014430 & & 7.77 & 10^{-6} & \\
6 & -1 & 401474 & & -1 & 401468 & & -1 & 4014757 & & -1 & 4014756 & & 1.35 & 10^{-6} & \\
\end{array}
\] 
\label{tbl:MPM_DMRG}
\end{table}
\acknowledgments
This work was supported by an award of the ANU STAC at the National Facility of the 
Australian Partnership for Advanced Computing.  Thanks to T. Nishino for useful discussions.

\end{document}